# Polyelectrolyte assisted charge titration spectrometry: applications to latex and oxide nanoparticles

F. Mousseau[1*], L. Vitorazi[1], L. Herrmann[1], S. Mornet[2] and J.-F. Berret[1*]

[1]Matière et Systèmes Complexes, UMR 7057 CNRS Université Denis Diderot Paris-VII, Bâtiment Condorcet, 10 rue Alice Domon et Léonie Duquet, 75205 Paris, France.
[2]Institut de Chimie de la Matière Condensée de Bordeaux, UPR CNRS 9048, Université Bordeaux 1, 87 Avenue du Docteur A. Schweitzer, Pessac cedex F-33608, France.

**Abstract:**
The electrostatic charge density of particles is of paramount importance for the control of the dispersion stability. Conventional methods use potentiometric, conductometric or turbidity titration but require large amount of samples. Here we report a simple and cost-effective method called polyelectrolyte assisted charge titration spectrometry or PACTS. The technique takes advantage of the propensity of oppositely charged polymers and particles to assemble upon mixing, leading to aggregation or phase separation. The mixed dispersions exhibit a maximum in light scattering as a function of the volumetric ratio $X$, and the peak position $X_{Max}$ is linked to the particle charge density according to $\sigma \sim D_0 X_{Max}$ where $D_0$ is the particle diameter. The PACTS is successfully applied to organic latex, aluminum and silicon oxide particles of positive or negative charge using poly(diallyldimethylammonium chloride) and poly(sodium 4-styrenesulfonate). The protocol is also optimized with respect to important parameters such as pH and concentration, and to the polyelectrolyte molecular weight. The advantages of the PACTS technique are that it requires minute amounts of sample and that it is suitable to a broad variety of charged nano-objects.

**Keywords**: charge density – nanoparticles - light scattering – polyelectrolyte complex

# I – Introduction

Electrostatic Coulomb forces are ubiquitous in soft condensed matter [1, 2]. Interaction pair potentials created by elementary charges of the same sign at interfaces or along macromolecules are long-range and repulsive. These interactions depend on physico-chemical parameters, such as the dielectric constant of the continuous phase, the solute concentration, the pH, the ionic strength and the temperature. Electrostatic forces between like-charged systems are especially relevant to insure repulsion between colloidal objects. At low ionic strength, electrostatic repulsions are for instance sufficient to induce long-range ordering and colloidal crystal phases [3]. In aqueous dispersions, the ionizable groups at the colloid surface exert Coulombic forces towards the counterions, leading to their condensation and to the formation of the electrical double layer [1]. The counterion condensation occurs for charge density $\sigma$ above a certain threshold [4]:

$$\sigma > \frac{2}{\pi D \ell_B} \quad (1)$$

where $D$ is the diameter of the colloid and $\ell_B$ is the Bjerrum length (0.7 nm in water). As a result, particles in solution satisfying Eq. 1 behave as if they would have an effective charge $Z_{eff}$ different from its structural charge $Z_{str}$ [4, 5]. The counterion condensation for colloids bears strong similarities with that derived for polymers and known as the Oosawa-Manning condensation [6]. Following Belloni [4] the effective charge can be approximated by the expression:

$$Z_{eff} = \frac{2D}{\ell_B} \quad (2)$$

For a particle of charge density $\sigma$ = $+1e$ nm$^{-2}$ and a diameter of 10 nm, the effective charge $Z_{eff}$ = $+28e$ represents around 10% of the total ionizable groups ($Z_{str}$ = $+314e$). As the condensed counterions are firmly attached to the surface and move with the colloid during diffusion, experiments such as electrophoretic mobility measurements [7-12] or small-angle scattering experiments in the concentrated regime [13-15] enable access to the effective charge only.

Potentiometric, turbidity or colloid titration techniques, as well as conductometry are commonly used to determine the structural charges of colloids. Potentiometric or acid-base titration coupled with conductometry was successfully applied to microgels [16], polymer micelles [17] and metal oxide nanoparticles [18-21]. However the technique requires large sample quantity, which depending on the particle synthesis is not always possible. To determine the charge density of iron oxide particles for instance, Lucas *et al.* used potentio-conductometric titration with dispersions containing 5-10 g of iron oxide dry matter [18]. Colloid titration is another method that was introduced by Terayama some 50 years ago [22]. This technique was applied to titrate ion-containing polymers in aqueous solutions. Colloid titration is based on the reaction between oppositely charged polyelectrolytes in presence of a small amount of dye molecules that serves as an indicator of the endpoint reaction between the two polymers [23, 24].

Initially developed to study protein complexes, isothermal titration calorimetry (ITC) has gained interest in the field of physical chemistry. ITC was also used to survey the condensation of DNA with multivalent counterions [25] and with oligo- and polycations [26-28] and more recently the complexation between polymers, proteins or surfactants. Despite a large number of studies [29-39] the interpretation of the thermograms remains a challenge, as the heat exchanges during titration exhibit rich and numerous features not always accounted for by existing models [33, 40]. More recently, ultrafast laser spectroscopy coupled to the generation of the second harmonics or resistive pulse sensing techniques were proposed to measure the surface charge of particles in the nanometer range [41, 42].

The technique examined here borrows its principle from methods developed for enzymatic activity measurements [43] and from turbidity titration [44, 45]. Earlier reports suggest that the maximum of absorbance or turbidity associated with the precipitation is related to the end-point reaction, and allows an estimation of the charge density of the titrated colloids. The method applied here differs from turbidity titration in two aspects: *i)* light scattering is preferred to UV-spectrometry and *ii)* the mixed polymer and particle dispersions are prepared by direct mixing instead of step-wise addition. With newly developed photon counters, light scattering is a highly sensitive technique, which can detect colloidal diffusion down to extremely low concentration. During the last decade, light scattering has been applied to investigate the complexation of oppositely charged species, including surfactant, polymers, phospholipids and proteins [12, 46-53]. In most studies however, emphasis was put on the structures that formed and not on the reaction stoichiometry [54]. Here we develop a simple protocol based on the use of light scattering and on the complexation property of particles with oppositely charged polymers. The technique is assessed on different types of organic and inorganic nanoparticles in the 50 nm range, either positive or negative and it is shown that the structural charge and charge density can be determined with minute amount of sample. The technique was dubbed Polyelectrolyte Assisted Charge Titration Spectrometry subsequently abbreviated as PACTS.

## II – Materials and Methods

### II.1 - Nanoparticles

Latex particles functionalized with carboxylate or amidine surface groups were acquired from Molecular Probes (concentration 40 g L$^{-1}$). The dispersion pH was adjusted at pH9.7 and pH6 by addition of sodium hydroxide and hydrochloric acid, respectively. The particles were characterized by light scattering and transmission electron microscopy, yielding $D_H$ = 39 nm and 56 nm and $D_0$ = 30 and 34 nm. Negative silica particles (CLX®, Sigma Aldrich) were diluted from 450 to 50 g L$^{-1}$ by DI-water. Particles were dialyzed for two days against DI-water at pH 9. Positive silica particles ($D_H$ = 60 nm) were synthetized using the Stöber synthesis route [55-57]. Silica seeds were first prepared and grown to increase the particle size. Functionalization by amine groups was then performed, resulting in a positively charged coating [56]. Aminated silica synthesized at 40 g L$^{-1}$ were diluted with DI-water and the pH was adjusted to pH 5 with hydrochloric acid. Aluminum oxide nanoparticle powder (Disperal®, SASOL) was dissolved in a nitric acid solution (0.4 wt. % in deionized water) at the concentration of 10 g L$^{-1}$ and sonicated for one hour. For the PACTS experiments, the dispersions were further diluted to 0.1 g L$^{-1}$ and the dispersion pH was adjusted to pH 4 with sodium hydroxide. In this pH condition, the nanoalumina are positively charged ($D_H$ = 64 nm) [58].

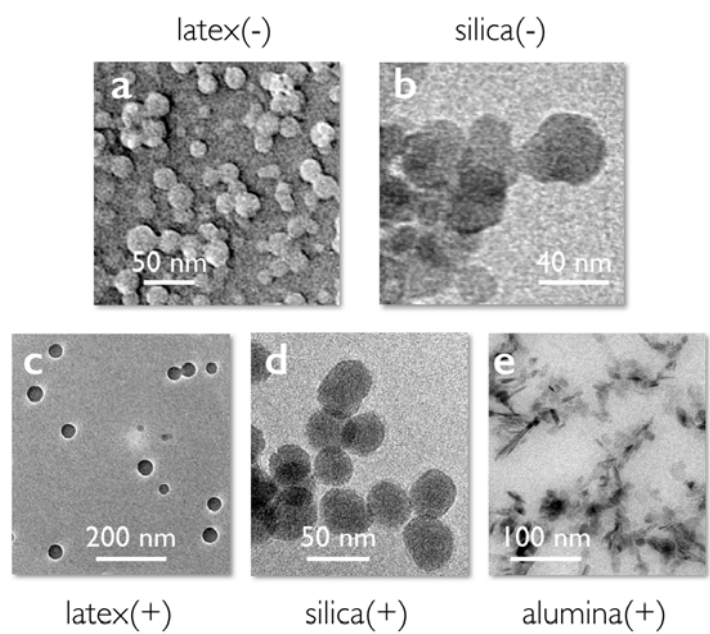

*Figure 1:* Transmission electronic microscopy images of latex (a,c), silica (b,d) and alumina (e) particles used in this study.

| Nano-particle | Chemical composition | Function-nalization | ρ (g cm$^{-3}$) | working pH | $D_H$ (nm) | $D_0$ (nm) | s |
|---|---|---|---|---|---|---|---|
| Latex (-) | Polystyrene | Carboxylate | 1.05 | 9.7 | 39 | 30 | 0.15 |
| Latex (+) | Polystyrene | Amidine | 1.05 | 6 | 56 | 34 | 0.15 |
| Silica (-) | SiO$_2$ | / | 2.3 | 9 | 34 | 20 | 0.20 |
| Silica (+) | SiO$_2$ | Amine | 1.9 | 5 | 60 | 42 | 0.11 |
| Alumina (+) | Al$_2$O$_3$ | / | 3.0 | 4 | 64 | 40 | 0.30 |

*Table I*: Nanoparticles studied in the present work. Particle characteristics are the mass density (ρ), the pH at which the experiments are done, the hydrodynamic diameter ($D_H$), the geometric diameter ($D_0$) and the dispersity s obtained from transmission electron microscopy.

Fig. 1 displays images of the particles obtained by transmission electron microscopy (TEM). Except for Al$_2$O$_3$ which have the form of irregular platelets of average dimensions 40 nm long and 10 nm thick, all other particles are spherical. In Table I, the hydrodynamic diameters $D_H$ are found slightly higher than the geometric diameters $D_0$ found by TEM. With light scattering, the largest particles contribute predominantly to the scattering intensity and determine the value for the measured $D_H$. The dispersity s defined as the ratio between the standard deviation and the average diameter are shown in Table I (see Supplementary Fig. 1 for size distributions).

### II.2 - Polymers

Poly(diallyldimethylammonium chloride) (PDADMAC) of molecular weight $M_w$ = 13.4 and 26.8 kDa was purchased from Polysciences Europe and from Sigma Aldrich respectively. The degree of polymerization was determined from the number-average molecular weight $M_n$ by size exclusion chromatography (SEC) and found to be 31 and 50 with respective dispersities of Đ = 2.7 and 3.5 [59]. As shown in Supplementary Fig. 2, SEC combined light-scattering detection using Novema columns reveals a double peak distribution for PDADMAC26.8k, one associated with a molecular weight around 30 kDa and one attributed to longer chains or polymer aggregates. Poly(sodium 4-styrenesulfonate) (PSS) of molecular weight $M_w$ = 8.0 kDa and 59.7 kDa were obtained from SRA Instruments and Sigma Aldrich. The degrees of polymerization were found at 32 and 137 by SEC, with respective dispersities Đ = 1.2 and 2.1. PDADMAC and PSS polymers were selected for titration because their ionization state does not depend on pH. Experiments were also carried out using poly(acrylic acid) (PAA, Sigma Aldrich) for comparison. Its degree of polymerization was obtained from SEC and found at 32, with a dispersity of 1.8. PAA is a weak polyelectrolyte characterized by a pKa of 5.5. The polymer characteristics are summarized in Table II and in Supplementary Fig. 2 [59]. The PDADMAC, PSS and PAA repetitive unit molar masses ($m_n$) are 161.5, 206.1 and 94.0 g mol$^{-1}$ respectively. Sodium acetate (CH$_3$COO$^-$, Na$^+$; 3H$_2$O), acetic acid, nitric acid, hydrochloric acid and sodium hydroxide were purchased from Sigma-Aldrich. Water was deionized with a Millipore DI-Water system. All the products were used without further purification. Stock solutions are diluted with DI-water to 1 or 0.1 g L$^{-1}$ and pH is adjusted with hydrochloric acid or with sodium hydroxide, depending on the particles to be titrated.

| Polymer | Provider | $M_n$ (kDa) | $M_w$ (kDa) | DP | Đ | Charge | Polyelectrolyte type |
|---|---|---|---|---|---|---|---|
| PDADMAC | Sigma Aldrich | 7.6 | 26.8 | 50 | 3.5 | + | Strong |
| PDADMAC | Polysciences Europe | 4.9 | 13.4 | 31 | 2.7 | + | Strong |
| PSS | Sigma Aldrich | 28.2 | 59.7 | 137 | 2.1 | - | Strong |
| PSS | SRA Instruments | 6.7 | 8.0 | 32 | 1.2 | - | Strong |
| PAA | Sigma Aldrich | 3.0 | 5.4 | 32 | 1.8 | - | Weak |

**Table II**: *Polyelectrolytes investigated in this work. Polymer characteristics are the number-average molecular weight ($M_n$), molecular weight ($M_w$), the degree of polymerization (DP), the dispersity (Đ) and the nature of the polyelectrolyte [59].*

### II.3 - Mixing protocols

For PSS, PAA and PDADMAC polymers, 500 µL batches were prepared in the same conditions of $pH$ and concentration (0.1, 1 or 10 g L$^{-1}$). Solutions were mixed at different charge ratios $Z_{-/+} = [-]/[+]$ where $[-]$ and $[+]$ denote the molar charge concentrations. This procedure was preferred to titration experiments because it allowed exploring a broad range in mixing conditions ($Z_{-/+} = 10^{-4} - 10^3$), while keeping the total concentration in the dilute regime [46, 60, 61]. Interactions between polymers occurred rapidly upon mixing, *i.e.* within a few seconds, and the dispersions were then studied by light scattering. The complexation of nanoparticles with oppositely charged polymers was investigated using a similar protocol. Particle concentration was adjusted so that the dispersion scattering intensity did not exceed $5 \times 10^5$ kcps, corresponding to a Rayleigh ratio $\mathcal{R}$ = $5 \times 10^{-3}$ cm$^{-1}$. Polymer and nanoparticle batches (500 µL) were prepared in the same $pH$ and concentration conditions (between 0.1 g L$^{-1}$ and 1 g L$^{-1}$). The solutions were mixed at different volumetric ratios $X$, where $X = V_{Pol}/V_{NP}$ and $V_{Pol}$ and $V_{NP}$ are the volumes of the polymer and particle solutions respectively. Because the stock solution concentrations are identical, the volumetric ratio $X$ is equivalent to the mass ratio between constituents.

### II.4 - Transmission electron microscopy
TEM imaging was performed with a Tecnai 12 operating at 80 kV equipped with a 1K×1K Keen View camera. Drops of suspensions (20 µL at 0.01 g L$^{-1}$ in DI-water) were deposited on holey-carbon coated 300 mesh copper grids (Neyco). Grids were let to dry over night at room temperature.

### II.5 – Isothermal Titration Calorimetry
Isothermal titration calorimetry (ITC) was performed using a Microcal VP-ITC calorimeter (Northampton, MA) with cell of 1.464 mL, working at 25 °C and agitation speed of 307 rpm. The syringe and the measuring cell were filled with degased solutions of PDADMAC, and PSS at the same pH. Water was also degased and filled the reference cell. Typical charge concentrations were 10 mM in the syringe and 1 mM in the measuring chamber. The titration consisted in a preliminary 2 µL injection, followed by 28 injections of 10 µL at 10 min intervals. A typical ITC experiment includes the thermogram (i.e. the differential power provided by the calorimeter to keep the temperature of cell and reference identical) and binding isotherm. Control experiments were carried out to determine the enthalpies associated to dilution. These behaviors were later subtracted to obtain the neat binding heat.

### II.6 – Static and Dynamic Light scattering
Light scattering measurements were carried out using a NanoZS Zetasizer (Malvern Instruments). In the light scattering experiment (detection angle at 173°), the hydrodynamic diameter $D_H$ and the scattered intensity $I_S$ were measured. The Rayleigh ratio $\mathcal{R}$ was derived from the intensity according to the relationship: $\mathcal{R} = (I_S - I_w)n_0^2\mathcal{R}_T/I_T n_T^2$ where $I_w$ and $I_T$ are the water and toluene scattering intensities respectively, $n_0$ = 1.333 and $n_T$ = 1.497 the solution and toluene refractive indexes, and $\mathcal{R}_T$ the toluene Rayleigh ratio at $\lambda$ = 633 nm ($\mathcal{R}_T = 1.352 \times 10^{-5}\ cm^{-1}$). The second-order autocorrelation function is analyzed using the cumulant and CONTIN algorithms to determine the average diffusion coefficient $D_C$ of the scatterers. Hydrodynamic diameter is then calculated according to the Stokes-Einstein relation, $D_H = k_B T/3\pi\eta D_C$, where $k_B$ is the Boltzmann constant, $T$ the temperature and $\eta$ the solvent viscosity. Measurements were performed in triplicate at 25 °C after an equilibration time of 120 s.

### II.7 – Electrophoretic mobility and zeta potential
Laser Doppler velocimetry using the phase analysis light scattering mode and detection at an angle of 16° was used to carry out the electrokinetic measurements of electrophoretic mobility and zeta potential with the Zetasizer Nano ZS equipment (Malvern Instruments, UK). Zeta potential was measured after a 2 min equilibration at 25 °C.

### II.8 - Optical microscopy
Phase-contrast images were acquired on an IX73 inverted microscope (Olympus) equipped with an 60× objectives. PSS8.0k and PDADMAC13.4k were prepared in MilliQ water at 20 g L$^{-1}$. Solutions were diluted to 0.1 g L$^{-1}$ with DI-water or 100 mM NaCl solution electrolyte. The polymer dispersions were mixed at $Z = 1$ and after 10 min, the mixture was diluted by a factor 10 for observation. 30 µl of dispersion were deposited on a glass plate and sealed into a Gene Frame® (Abgene/Advanced Biotech) dual adhesive system. An EXi Blue camera (QImaging) and Metaview software (Universal Imaging Inc.) were used as the acquisition system.

## III - Results
### III.1 – Assessment of the PACTS technique using ion-containing polymers
The PACTS technique was first assessed using oppositely charged polymers. Poly(diallyldimethylammonium chloride) and poly(sodium 4-styrenesulfonate) of molecular weight $M_w$ = 13.4 and 8.0 kDa with similar degrees of polymerization ($DP$ = 31 and 32 respectively) were considered. In this study, isothermal titration calorimetry was used to determine the stoichiometry of the inter-polyelectrolyte reaction, and to set up a reference for PACTS. In addition to the charge stoichiometry, ITC also provides the binding enthalpy and reaction binding constant. Figs. 2a and 2b display the thermogram and the binding isotherm obtained for PSS/PDADMAC, respectively. Here, the PSS dispersion at molar charge concentration 20 mM was added stepwise to a PDADMAC solution containing 2 mM of positive charges. Throughout the process, the enthalpy exhibits a sigmoidal decrease with increasing charge ratio and is associated with an exothermic reaction. Above $Z_{-/+}$ = 1, heat exchanges close to zero indicate that the titration is completed. Experiments performed with PSS and PDADMAC of different molecular weights provide similar thermograms (see Supplementary Fig. 3). The data of Fig. 2 also confirm the ITC results obtained by Bucur and coworkers on the same polymers at a slightly different ionic strength (0.3 M NaCl) [29].

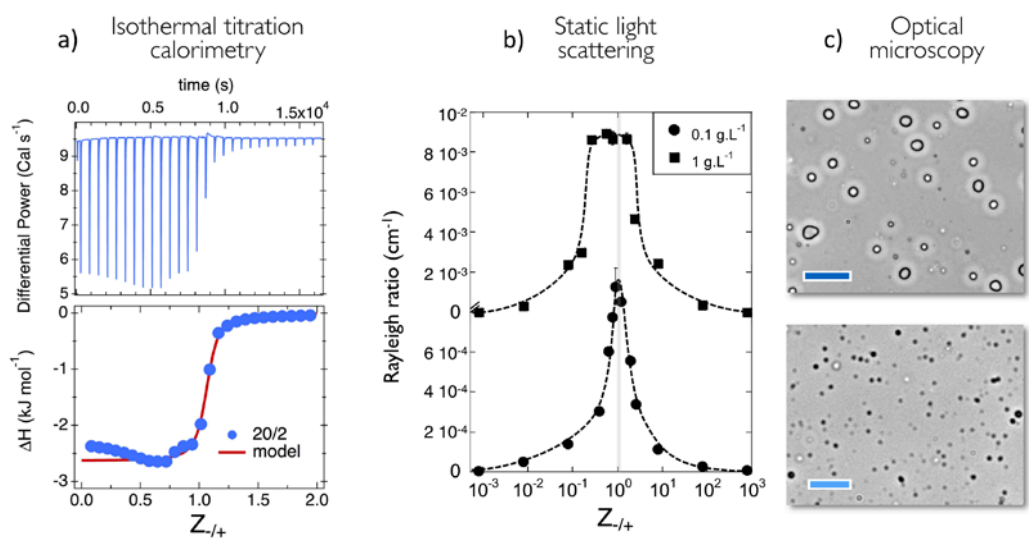

*Figure 2:* a) Thermogram (upper panel) and binding isotherm (lower panel) obtained from micro-calorimetry titration of poly(diallyldimethylammonium chloride) (PDADMAC $M_w$ = 13.4 kDa) by stepwise addition of poly(sodium 4-styrenesulfonate) (PSS, $M_w$ = 8.0 kDa.) The molar charge concentrations are 20 mM and 2 mM respectively, and the temperature is fixed at T = 25 °C. The continuous curve in red arises from best fit calculations using the Multiple Non-interacting Sites (MNIS) model. The stoichiometry coefficient derived from ITC is n = 1.1 ± 0.1. b) Rayleigh ratios obtained from PDADMAC/PSS mixed dispersions formulated by direct mixing. The scattering intensity exhibits a maximum at the 1:1 charge stoichiometry. c) Bright field optical microscopy images of the PDADMAC/PSS coacervate phase prepared without added salt (upper panel) and at the ionic strength of 100 mM. The dark (light) blue bars are 20 (5) µm.

The ITC data on PSS/PDADMAC were analyzed according to the Multiple Non-interacting Sites (MNIS) model [33, 40, 62]. The MNIS model assumes that the macromolecules to be titrated have several anchoring sites, and that the binding probability is independent on the occupation rate of other sites. This simplified model was found to work well for a broad range of colloidal systems [33, 39]. The reaction between macromolecules is associated with heat exchange that is proportional to the amount of binding events and characterized by the enthalpy $\Delta H_b$, the binding constant $K_b$ and by a reaction stoichiometry $n$. Fig. 2a displays the ITC data together with the fitting curves obtained from the MNIS model. The values retrieved are the binding enthalpy $\Delta H_b$ = -2.6 kJ mol$^{-1}$, the free energy $\Delta G = -RTLnK_b$ = -30.4 kJ mol$^{-1}$ and the entropy contribution to the reaction $T\Delta S = \Delta H_b - \Delta G$ = + 27.8 kJ mol$^{-1}$. The associated binding constant is $K_b = 2.0 \times 10^5$ M$^{-1}$ and the stoichiometry coefficient $n$ = 1.1 ± 0.1. For PDADMAC13.4k and PDADMAC26.8k titrated alternatively with PSS8.0k and PSS59.7k, stoichiometry coefficients close to unity were also obtained (Supplementary Fig. 3). The large value of the binding constant indicates a strong affinity of styrene sulfonate for diallyldimethylammonium, and a stoichiometry around 1 that the complexation occurs through charge neutralization. The entropy contribution to the reaction, $-T\Delta S$ is around 10 times larger than the binding enthalpy, demonstrating that the process is driven by the entropy and dominated by the release of the sodium and chloride counterions [25, 54].

The PACTS technique was applied to PSS/PDADMAC following the direct mixing protocol described in Section II.3. The volumes of the stock solutions used for mixing were adjusted to cover a range in charge ratio between $10^{-4}$ and $10^3$. After mixing, the dispersions were stirred rapidly, let to equilibrate for 5 minutes and the scattered intensity and hydrodynamic diameter were measured in triplicate. The light scattering experiments were repeated a day after and showed the same features (data not shown). Fig. 2b displays the Rayleigh ratio obtained for 0.1 and 1 g L$^{-1}$ PSS/PDADMAC mixtures. Both datasets exhibit a maximum around the 1:1 charge stoichiometry. At 1 g L$^{-1}$, the scattering peak is broad and exhibits a flat plateau between $Z_{-/+}$ = 0.5 and 2. At 0.1 g L$^{-1}$ in contrast, the sharp maximum is observed and allows an accurate determination of its location, here at $Z_{Max}$ = 1.2 ± 0.1. Such a scattering feature on mixing oppositely charge species was observed for various systems, including synthetic and biological polymers, phospholipid vesicles and surfactants [19, 31, 32, 47, 48, 51, 52, 63]. It is interpreted in terms of complexation, charge neutralization and the formation of structures or phase much larger the initial components. In this scenario, the coacervate particles formed at equivalence ($[+] = [-]$) have a zero surface charge and grow rapidly in size. In off-stoichiometric mixtures on the other hand, the coacervate particles are charged and repel each other, preventing them from growing. The $Z_{max}$-value is also in good agreement with the stoichiometry found in ITC on the same compounds. We have found that stoichiometric PSS/PDADMAC mixtures without added salt form a viscous coacervate phase that sedimented rapidly [63, 64]. Observed between glass slides, the coacervate appears as 5 µm droplets (Fig. 2c, top panel). At 0.1 M of added salt, the phase separation persists and gives rise to droplets in the micron range (Fig. 2c, lower panel). These polymer phases are similar to those found by Priftis *et al.* on branched poly(ethyleneimine) complexed with linear poly(glutamic acid) [51, 65], indicating that electrostatics driven reactions between charged polymers share general features as far as the structure, the phase or the thermodynamics are concerned [32]. The main result of this study is that the PACTS technique allows an accurate determination of the charge stoichiometry.

### III.2 – Application of the PACTS technique to nanoparticles and polymers

Polymer/nanoparticle dispersions were formulated by mixing stock solutions at different volumetric ratios between $10^{-4}$ and $10^3$. Fig. 3a, 3b and 3c show the Rayleigh ratio $\mathcal{R}(X)$, hydrodynamic diameter $D_H(X)$ and zeta potential $\zeta(X)$ respectively for latex/poly(sodium 4-styrenesulfonate) mixtures. At 0.1 g L$^{-1}$, the mixed dispersions are dilute and the scattering intensity is proportional to the concentration and to the molecular weight of the scatterers. The continuous line in Fig. 3a is calculated assuming that polymers and particles do not interact, and that the Rayleigh ratio is the sum of the respective Rayleigh ratios weighted by their actual concentrations [60]:

$$\mathcal{R}_0(X) = \frac{\mathcal{R}_{NP} + \mathcal{R}_{Pol} X}{1 + X} \qquad (3)$$

where $\mathcal{R}_{NP}$ and $\mathcal{R}_{Pol}$ are the Rayleigh ratios of the nanoparticle and polymer dispersions, respectively. In Figs. 3a, $\mathcal{R}(X)$ is found to be higher than the predictions for non-interacting species, and it exhibits a marked maximum at the critical value $X_{Max}$ = 0.020 ± 0.003..

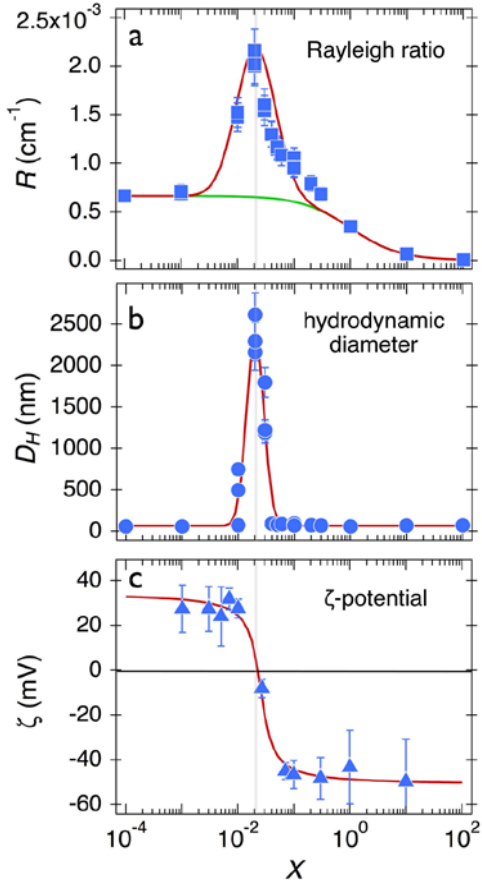

*Figure 3: Rayleigh ratio $\mathcal{R}(X)$ (a), hydrodynamic diameter $D_H(X)$ (b) and zeta potential $\zeta(X)$ (c) measured for positively charged 34 nm latex particles complexed with poly(sodium 4-styrenesulfonate) chains of molecular weight $M_w$ = 8.0 kDa at concentration c = 0.1 g L$^{-1}$. The continuous line in green (a) is calculated according to Eq. 3. The position of $\mathcal{R}(X)$ and $D_H(X)$-maxima coincides with charge neutralization and a zero zeta potential. Continuous lines in red are guides for the eyes.*

The observation of a maximum suggests that polymer/particle aggregates are formed, and that it occurs at a defined stoichiometry. The hydrodynamic diameter $D_H(X)$ in Fig. 3b confirms this result. The $D_H$-maximum reaches a value over a micron at the same $X_{Max}$ as the intensity. Fig. 3c shows the zeta potential $\zeta(X)$ as a function of the mixing ratio. The point of zero charge, characterized by $\zeta(X) = 0$ is obtained for $X = X_{Max}$, indicating that the reaction stoichiometry is associated with the neutralization of the electrostatic charges [31, 52, 66]Zeta measurements using higher molecular weight poly(sodium 4-styrenesulfonate) (Mw = 59.7 kDa) confirm this result (Supplementary Fig. 4). Our goal here is to show that under appropriate conditions, the approach can be exploited to determine the structural charges of particulate nanosystems.

**III.3 - Charge density calculation and comparison with standard methods**

Fig. 4 displays the Rayleigh ratio and the hydrodynamic diameter for latex (a), silica (b,c) alumina (d). The negative latex and silica were complexed with PDADMAC13.4 k, whereas the positively charged particles (silica and alumina) were associated with the 8 kDa PSS. The features disclosed in Fig. 3 were again observed: both $\mathcal{R}(X)$- and $D_H(X)$-data exhibit sharp maxima. The positions of these maxima coincide and allow an accurate determination of $X_{Max}$. The values for the mixing ratios at maximum are 0.0023 ± 0.0003 for carboxylate coated latex, 0.010 ± 0.002 for negative silica, 0.016 ± 0.004 for amine coated silica and 0.10 ± 0.01 for alumina. Assuming that the peak position coincides with the charge neutralization, the structural charge $Z_{Str}$ is obtained from:

$$Z_{Str} = X_{Max} \frac{M_n^{Np}}{m_n} \quad (4)$$

where $M_n^{Np}$ and $m_n$ denote the number-averaged molecular weights of the particles and of the repetitive unit of the polymer used for titration, respectively. Taking into account the particle dispersity $s$ (Table I), the charge density reads:

$$\sigma = \frac{\mathcal{N}_A \rho D_0 \, exp(2.5s^2)}{6m_n} X_{Max} \quad (5)$$

In Eq. 5, $\rho$ is the particle mass density and $\mathcal{N}_A$ the Avogadro number. For log-normal distribution of median diameter $D_0$ and dispersity $s$, the i$^{th}$-moment is given by the expression $<D^i> = D_0^i exp\,(i^2s^2/2)$. For particles of uniform size ($s \sim 0$), the exponential term in the numerator of Eq. 5 is close to 1, and the charge density is directly proportional to the particle diameter and the scattering maximum position observed by PACTS.

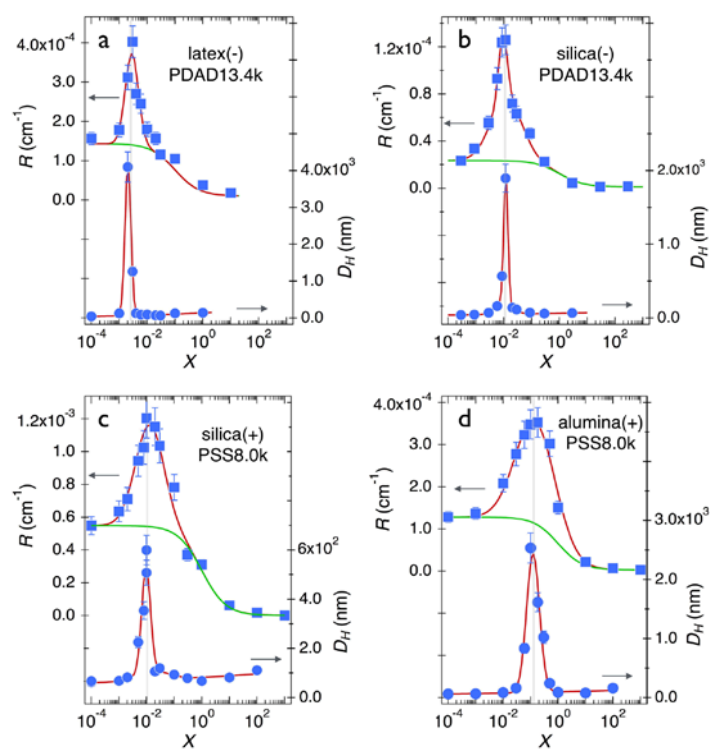

**Figure 4:** Rayleigh ratio $\mathcal{R}(X)$ and hydrodynamic diameter $D_H(X)$ obtained from PACTS experiments performed with latex (a), silica (b and c) and alumina (d) nanoparticles. The legends are similar to those of Fig. 3. Continuous lines in red are guides for the eyes. PACTS was performed at concentrations $c = 1$ g L$^{-1}$ in a), $c = 0.1$ g L$^{-1}$ in b), $c = 0.5$ g L$^{-1}$ in c) and $c = 0.1$ g L$^{-1}$ in d).

| Nanoparticles(sign) | nominal charge density $\sigma$ (nm$^{-2}$) | Mixing ratio at maximum scattering $X_{Max}$ | charge density from PACTS $\sigma$ (nm$^{-2}$) |
|---|---|---|---|
| Latex(-) | -0.05 | 0.0023 ± 0.003 | -0.048 ± 0.006 |
| Latex(+) | +0.21 | 0.020 ± 0.003 | +0.33 ± 0.06 |
| Silica (-) | -0.24 | 0.010 ± 0.002 | -0.31 ± 0.07 |
| Silica (+) | n.d. | 0.016 ± 0.004 | +0.62 ± 0.16 |
| Alumina(+) | +5.9 | 0.10 ± 0.01 | +7.3 ± 0.7 |

**Table III:** Charge densities $\sigma$ of the particles studied in this work determined by conventional titration methods and by PACTS. $X_{Max}$ denotes the volumetric mixing ratio at which the scattering intensity and hydrodynamic diameter are maximum. The charge density from PACTS was calculated from Eq. 5 and particle characteristics of Table I.

Results are summarized in Table III, and the charge densities obtained from Eq. 5 are compared with those retrieved from conventional methods. The $\sigma$-values for carboxylate and amidine latex ($\sigma$ = -0.05$e$ nm$^{-2}$ and +0.21$e$ nm$^{-2}$ respectively) were provided by the supplier, whereas those of negative silica and alumina were measured using potentiometry and precipitation titration coupled to conductometry experiments. The positive silica particles were produced in small quantities and their charges could not be obtained by titration. As shown in Table III for alumina, latex and negative silica, PACTS provides charge densities in agreement with the other techniques. In view of the dispersity exhibited by the polymers and particles examined in this work, a 20 - 30% difference between PACTS and other techniques is noticeable. For positive silica, the value retrieved from PACTS amounts at $\sigma$ = +0.62$e$ nm$^{-2}$, a value in fair agreement with earlier determination [56, 57]. From the charge densities listed, it can also be verified that the inequality $\sigma > 2/\pi D \ell_B$ (Eq. 1) holds and that the condensation and double layer description apply for all particles considered [1]. In conclusion, it can be said that the PACTS is an accurate technique for measuring the structural charges of nanoparticles.

### III.4 – Optimizing PACTS

In this section, we explore the effects of physico-chemical parameters such as the concentration, the molecular weight and the polymer ionization state.

*Effect of concentrations and sensitivity:* As illustrated in Fig. 2 for PSS8.0k and PDAD13.4k, the initial concentration has a significant effect on the scattering peak shape. The plateau observed at 1 g L$^{-1}$ around the 1:1 charge stoichiometry arises from the high turbidity of the coacervate phase. As a result, the sample absorbs a noticeable part of the incoming and scattered light, and the recorded values of the Rayleigh ratios are biased. At 0.1 g L$^{-1}$, the hydrodynamic sizes are of the order of 200 nm, the turbidity is lowered, making measurements possible. For particles, PACTS measurements were made between 0.1 and 1 g L$^{-1}$ and gave Rayleigh ratios around 10$^{-3}$ cm$^{-1}$ at the peak maximum. These values are 1000 times larger than the minimum Rayleigh ratio detectable by light scattering spectrometers [49]. The concentration in a PACTS experiment could hence be reduced and still provides reliable data. One important advantage of PACTS over the techniques mentioned in introduction is that it requires very low amount of samples.

*Effect of polymer molecular weight:* The molecular weight of the titrating polymers plays also an important role. Fig. 5a displays the Rayleigh ratio as a function of $X$ for negative silica complexed with poly(diallyldimethylammonium chloride) of molecular weight $M_w$ = 13.4 kDa and 26.8 kDa. With increasing molecular weight, the scattering peak broadens and shifts toward higher mixing ratios, here from $X_{Max}$ = 0.01 to 0.02, leading to a doubling of the apparent charge density (-0.30e nm$^{-2}$ versus -0.61e nm$^{-2}$). Similarly, positive latex particles were associated with poly(sodium 4-styrenesulfonate) at $M_w$ = 8.0 kDa and $M_w$ = 59.7 kDa. Again, the scattering peak broadens and moves toward higher $X$-values (Fig. 5b). With PSS59.7k, the charge density deduced using Eq. 5 equals $\sigma$ = +2.6e nm$^{-2}$, that is seven times that obtained with PSS8.0k. These results suggest that the particle aggregation mechanism depends on the molecular weight of the titrating polymers. For short chains, it is assumed that the polymers adsorb at the particle surface and compensate the structural charges. As a result the destabilization occurs through the surface charge neutralization and electrostatic screening. Note here that the structural charges of the polymers ($Z_{str} \sim 30$) are much smaller than that of the particles ($Z_{str} = 1000 - 5000$). This asymmetry appears as an important criterion in regulating the adsorption and precipitation processes. With higher molecular weight polymers, polymers adsorb at the particle surface but form loops or bridges with neighboring particles. In this latter case, there is an excess of uncomplexed charges, which artificially increases the particle charge density.

*Effect of polyelectrolyte:* Fig. 5c compares the Rayleigh ratios $\mathcal{R}(X)$ using a weak (poly(sodium acrylate) 5.4 kDa) and a strong (PSS8.0k) polyelectrolyte of same degree of dispersity, together with the positively charged silica. Experiments were performed under conditions where the particles are colloidally stable, i.e. at pH5. Acid-base titration experiments on PAA 5.4 kDa have shown that at this pH, the chain ionization degree is 30%. Fig. 5c illustrates that switching from strong to weak poly(acid) shifts the scattering maximum position to higher values, here from $X_{Max}$ = 0.013 to 0.022. As a result, the charge density estimated from Eq. 5 is found to be higher with PAA ($\sigma$ = +1.85e nm$^{-2}$) as compared to that with PSS ($\sigma$ = +0.63e nm$^{-2}$). The reason for this discrepancy is that not all the acrylic acid monomers adsorbing at the interface are charged and complex with the silica surface charges. Using PDADMAC and partially charged PAA chains, it was shown recently that the complexation process itself modifies the degree of ionization of the weak poly(acid) chains and displaces the reaction stoichiometry [39, 67].

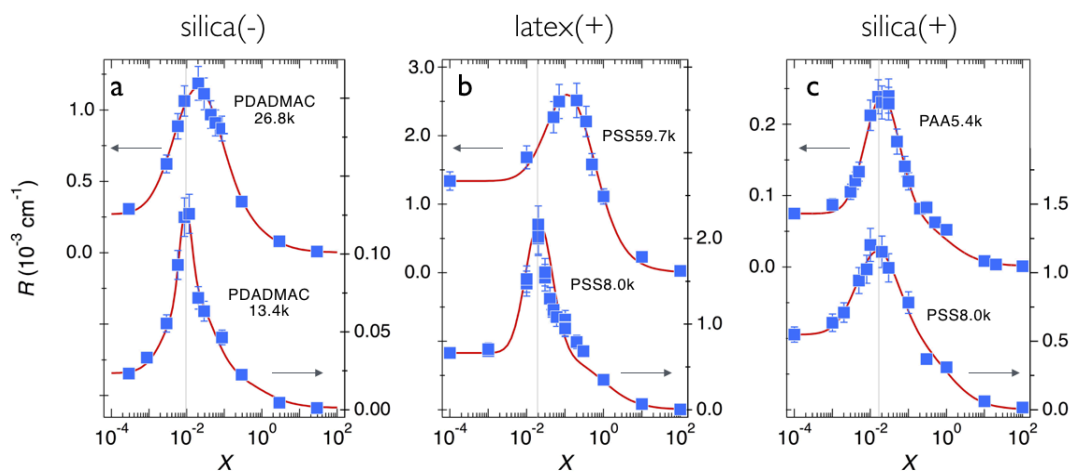

**Figure 5**: Rayleigh ratio $\mathcal{R}(X)$ and hydrodynamic diameter $D_H(X)$ obtained from PACTS experiments performed with negative silica (a), positive latex (b) and silica (c) nanoparticles. Except for silica(-)/PDADMAC26.8k that was measured at $c = 1$ g L$^{-1}$, all PACTS concentrations were $c = 0.1$ g L$^{-1}$. The legends are similar to those of Fig. 3. Continuous lines in red are guides for the eyes.

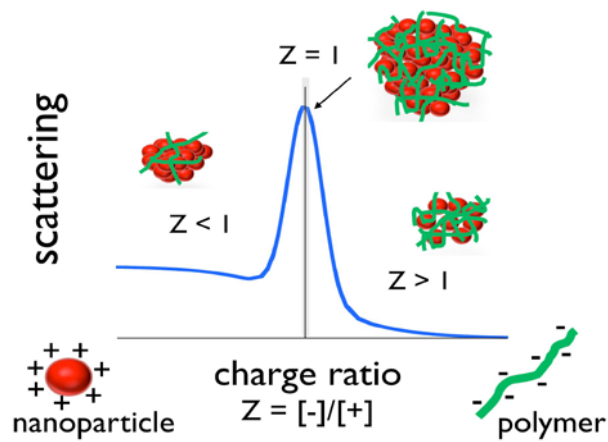

**Figure 6**: Schematic representation of the light scattering response to mixing oppositely charged polymers and particles. The position of the maximum scattering peak corresponds to the 1:1 charge stoichiometry.

## IV – Conclusion

The main result of this study is that polyelectrolyte assisted charge titration spectrometry technique is a rapid and effective method to estimate nanoparticle structural charges. Here we are taking advantage of the propensity of oppositely charged polymers and particles to assemble upon mixing. The designed complexation protocole makes use of low molecular weight ion-containing polymers as complexing agents. Upon mixing, the chains adsorb at the particle surface in an entropy-driven process and compensate the charge of the nanoparticles. The particles then aggregate via van der Waals interactions, resulting in the formation of micron size objects, or in some cases micro-phase separation. Both induced states are easily detectable by light scattering. The screening of the polymer/nonoparticle phase diagram by varying the volumetric ratio leads to a marked scattering peak (Fig. 6). The peak position on the $X$-axis provides the value of the charge density $\sigma$, assuming its size and dispersity are known. In this approach, the complexing polymers are of low molecular weight to avoid overcharging during the adsorption process. Overcharging can be associated to the formation of loops, dangling ends or to the bridging of distant particles. The use of high molecular weight polymers leads to a wrong estimation of the structural charges, as shown in Fig. 5. The protocol was further optimized with respect to pH, concentration and to the nature of the polymers. The technique used here is similar in its principle to the one reported by one of us a few years ago [60]. In this previous work, the complexation was carried out using double hydrophilic block copolymers instead of homopolyelectrolytes. The role of the neutral block was to reduce the interfacial tension between the particle aggregates and the solvent, and to stabilize the microstructure in the 100 nm range. With homopolyelectrolytes the situation is different, as the complexation and further growth of the aggregates are not hindered by the presence of neutral blocks, therefore leading to the formation of micron size colloids or coacervate droplets solely at the charge stoichiometry. One decisive advantage of PACTS is that it requires minute amounts of particles, typically 10 µg of dry matter, whereas conventional titration uses $10^3 - 10^4$ larger quantities [16-18, 21]. In conclusion, PACTS represents a quick and easy protocol that can be used to determine the structural charge density of nanoparticles, which is of critical importance for the stability and interaction of bulk dispersions.

## Acknowledgments

We thank Jean-Paul Chapel, Fabrice Cousin, Jacques Jestin, Evdokia Oikonomou, Olivier Sandre and Christophe Schatz for fruitful discussions. L.V. thanks CNPq (Conselho Nacional de Desenvolvimento Científico e Tecnológico) in Brazil for postdoctoral fellowship (Proc. N°. 210694/2013-0). We also thank Mélanie Legros from Institut Charles Sadron in Strasbourg, France for the size-exclusion chromatography measurements and molecular weight determination. ANR (Agence Nationale de la Recherche) and CGI (Commissariat à l'Investissement d'Avenir) are gratefully acknowledged for their financial support of this work through Labex SEAM (Science and Engineering for Advanced Materials and devices) ANR 11 LABX 086, ANR 11 IDEX 05 02. This research was supported in part by the Agence Nationale de la Recherche under the contract ANR-13-BS08-0015 (PANORAMA).

## References

[1] J.N. Israelachvili, Intermolecular and Surfaces Forces, 2nd ed., Academic Press, New York, 1992.
[2] D.F. Evans, K. Wennerström, The Colloidal Domain, Wiley-VCH, New York, 1999.
[3] P. Pieranski, Colloidal crystals, Contemporary Physics, 24 (1983) 25-73.
[4] B. L., Ionic Condensation and Charge Renormalization in Colloid Suspensions, Colloids Surf. A, 140 (1998) 227 - 243.
[5] S. Alexander, P.M. Chaikin, P. Grant, G.J. Morales, P. Pincus, D. Hone, Charge renormalization, osmotic-pressure, and bulk modulus of colloidal crystals - theory, J. Chem. Phys., 80 (1984) 5776-5781.
[6] F. Oosawa, Polyelectrolytes, Marcel Dekker, New York, N. Y., 1971.
[7] F. Strubbe, F. Beunis, K. Neyts, Determination of the effective charge of individual colloidal particles, J. Colloid Interface Sci., 301 (2006) 302-309.
[8] T. Trimaille, C. Pichot, A. Elaissari, H. Fessi, S. Briancon, T. Delair, Poly(D,L-lactic acid) nanoparticle preparation and colloidal characterization, Colloid Polym. Sci., 281 (2003) 1184-1190.
[9] J.P. Holmberg, E. Ahlberg, J. Bergenholtz, M. Hassellov, Z. Abbas, Surface charge and interfacial potential of titanium dioxide nanoparticles: Experimental and theoretical investigations, J. Colloid Interface Sci., 407 (2013) 168-176.
[10] K. Makino, H. Ohshima, Electrophoretic Mobility of a Colloidal Particle with Constant Surface Charge Density, Langmuir, 26 (2010) 18016-18019.
[11] A.V. Delgado, F. Gonzalez-Caballero, R.J. Hunter, L.K. Koopal, J. Lyklema, Measurement and interpretation of electrokinetic phenomena, J. Colloid Interface Sci., 309 (2007) 194-224.
[12] R. Doi, E. Kokufuta, On the Water Dispersibility of a 1:1 Stoichiometric Complex between a Cationic Nanogel and Linear Polyanion, Langmuir, 26 (2010) 13579-13589.
[13] B. D'aguanno, R. Klein, Integral-equation theory of polydisperse yukawa systems, Phys. Rev. A, 46 (1992) 7652-7656.
[14] N. Gorski, M. Gradzielski, H. Hoffmann, Mixtures of nonionic and ionic surfactants - the effect of counterion binding in mixtures of tetradecyldimethylamine oxide and tetradecyltrimethylammonium bromide, Langmuir, 10 (1994) 2594-2603.
[15] M. Quesada-Perez, J. Callejas-Fernandez, R. Hidalgo-Alvarez, Interaction potentials, structural

ordering and effective charges in dispersions of charged colloidal particles, Adv. Colloids Interface Sci., 95 (2002) 295-315.

[16] T. Hoare, R. Pelton, Titrametric characterization of pH-induced phase transitions in functionalized microgels, Langmuir, 22 (2006) 7342-7350.

[17] J. Rodriguez-Hernandez, J. Babin, B. Zappone, S. Lecommandoux, Preparation of shell cross-linked nano-objects from hybrid-peptide block copolymers, Biomacromolecules, 6 (2005) 2213-2220.

[18] I.T. Lucas, S. Durand-Vidal, E. Dubois, J. Chevalet, P. Turq, Surface charge density of maghemite nanoparticles: Role of electrostatics in the proton exchange, J. Phys. Chem. C, 111 (2007) 18568-18576.

[19] J. Fresnais, M. Yan, J. Courtois, T. Bostelmann, A. Bee, J.F. Berret, Poly(acrylic acid)-coated iron oxide nanoparticles: Quantitative evaluation of the coating properties and applications for the removal of a pollutant dye, J. Colloid Interface Sci., 395 (2013) 24-30.

[20] J. Sonnefeld, A. Gobel, W. Vogelsberger, Surface-charge density on spherical silica particles in aqueous alkali chloride solutions .1. Experimental results, Colloid Polym. Sci., 273 (1995) 926-931.

[21] F.A. Tourinho, A.F.C. Campos, R. Aquino, M. Lara, G.J. da Silva, J. Depeyrot, Surface charge density determination in electric double layered magnetic fluids, Braz. J. Phys., 32 (2002) 501-508.

[22] H. Terayama, Method of colloid titration (a new titration between polymer ions), J. Polym. Sci., 8 (1952) 243-253.

[23] T. Masadome, Determination of cationic polyelectrolytes using a photometric titration with crystal violet as a color indicator, Talanta, 59 (2003) 659-666.

[24] K. Ueno, K. Kina, Colloid titration - a rapid method for the determination of charged colloid, J. Chem. Educ., 62 (1985) 627-629.

[25] D. Matulis, I. Rouzina, V.A. Bloomfield, Thermodynamics of DNA binding and condensation: Isothermal titration calorimetry and electrostatic mechanism, J. Mol. Biol., 296 (2000) 1053-1063.

[26] W. Kim, Y. Yamasaki, K. Kataoka, Development of a fitting model suitable for the isothermal titration calorimetric curve of DNA with cationic ligands, J. Phys. Chem. B, 110 (2006) 10919-10925.

[27] W. Kim, Y. Yamasaki, W.-D. Jang, K. Kataoka, Thermodynamics of DNA Condensation Induced by Poly(ethylene glycol)-block-polylysine through Polyion Complex Micelle Formation, Biomacromolecules, 11 (2010) 1180-1186.

[28] N. Korolev, N.V. Berezhnoy, K.D. Eom, J.P. Tam, L. Nordenskiold, A universal description for the experimental behavior of salt-(in)dependent oligocation-induced DNA condensation (vol 37, pg 7137, 2009), Nucleic Acids Res., 40 (2012) 2807-2821.

[29] C.B. Bucur, Z. Sui, J.B. Schlenoff, Ideal mixing in polyelectrolyte complexes and multilayers: Entropy driven assembly, J. Am. Chem. Soc., 128 (2006) 13690-13691.

[30] T. Cedervall, I. Lynch, S. Lindman, T. Berggard, E. Thulin, H. Nilsson, K.A. Dawson, S. Linse, Understanding the nanoparticle-protein corona using methods to quantify exchange rates and affinities of proteins for nanoparticles, Proc. Natl. Acad. Sci. U. S. A., 104 (2007) 2050-2055.

[31] F. Loosli, P. Le Coustumer, S. Stoll, TiO2 nanoparticles aggregation and disaggregation in presence of alginate and Suwannee River humic acids. pH and concentration effects on nanoparticle stability, Water Research, 47 (2013) 6052-6063.

[32] J. Qin, D. Priftis, R. Farina, S.L. Perry, L. Leon, J. Whitmer, K. Hoffmann, M. Tirrell, J.J. de Pablo, Interfacial Tension of Polyelectrolyte Complex Coacervate Phases, ACS Macro Lett., 3 (2014) 565-568.

[33] J. Courtois, J.-F. Berret, Probing Oppositely Charged Surfactant and Copolymer Interactions by Isothermal Titration Microcalorimetry, Langmuir, 26 (2010) 11750-11758.

[34] L. Chiappisi, D. Li, N.J. Wagner, M. Gradzielski, An improved method for analyzing isothermal titration calorimetry data from oppositely charged surfactant polyelectrolyte mixtures, J. Chem. Thermodyn., 68 (2014) 48-52.

[35] I. Herrera, M.A. Winnik, Differential Binding Models for Isothermal Titration Calorimetry: Moving beyond the Wiseman Isotherm, J. Phys. Chem. B, 117 (2013) 8659-8672.

[36] S. Louguet, A.C. Kumar, N. Guidolin, G. Sigaud, E. Duguet, S. Lecommandoux, C. Schatz, Control of the PEO Chain Conformation on Nanoparticles by Adsorption of PEO-block-Poly(l-lysine) Copolymers and Its Significance on Colloidal Stability and Protein Repellency, Langmuir, 27 (2011) 12891-12901.

[37] D. Priftis, N. Laugel, M. Tirrell, Thermodynamic Characterization of Polypeptide Complex Coacervation, Langmuir, 28 (2012) 15947-15957.

[38] N. Welsch, A.L. Becker, J. Dzubiella, M. Ballauff, Core-shell microgels as "smart" carriers for enzymes, Soft Matter, 8 (2012) 1428-1436.

[39] L. Vitorazi, N. Ould-Moussa, S. Sekar, J. Fresnais, W. Loh, J.P. Chapel, J.F. Berret, Evidence of a two-step process and pathway dependency in the thermodynamics of poly(diallyldimethylammonium chloride)/poly(sodium acrylate) complexation, Soft Matter, 10 (2014) 9496-9505.

[40] T. Wiseman, S. Williston, J.F. Brandts, L.N. Lin, Rapid measurement of binding constants and heats of binding using a new titration calorimeter, Anal. Biochem., 179 (1989) 131-137.

[41] R.R. Kumal, T.E. Karam, L.H. Haber, Determination of the Surface Charge Density of Colloidal Gold Nanoparticles Using Second Harmonic Generation, J. Phys. Chem. C, 119 (2015) 16200-16207.
[42] R. Vogel, W. Anderson, J. Eldridge, B. Glossop, G. Willmott, A Variable Pressure Method for Characterizing Nanoparticle Surface Charge Using Pore Sensors, Anal. Chem., 84 (2012) 3125-3131.
[43] P. Job, Studies on the formation of complex minerals in solution and on their stability, Annales De Chimie France, 9 (1928) 113-203.
[44] H. Dautzenberg, Polyelectrolyte complex formation in highly aggregating systems. 1. Effect of salt: Polyelectrolyte complex formation in the presence of NaCl, Macromolecules, 30 (1997) 7810-7815.
[45] J.M. Lambert, Volumetric analysis of colloidal electrolytes by turbidity titration, J. Colloid Sci., 2 (1947) 479-493.
[46] P. Herve, M. Destarac, J.-F. Berret, J. Lal, J. Oberdisse, I. Grillo, Novel core-shell structure for colloids made of neutral/polyelectrolyte diblock copolymers and oppositely charged surfactants, Europhys. Lett., 58 (2002) 912-918.
[47] L. Leclercq, M. Boustta, M. Vert, A physico-chemical approach of polyanion-polycation interactions aimed at better understanding the in vivo behaviour of polyelectrolyte-based drug delivery and gene transfection, J. Drug Target., 11 (2003) 129-138.
[48] S. Sennato, F. Bordi, C. Cametti, Correlated adsorption of polyelectrolytes in the "charge inversion" of colloidal particles, Europhys. Lett., 68 (2004) 296-302.
[49] L. Qi, J.P. Chapel, J.C. Castaing, J. Fresnais, J.-F. Berret, Organic versus hybrid coacervate complexes: co-assembly and adsorption properties, Soft Matter, 4 (2008) 577-585.
[50] L. Qi, J. Fresnais, J.-F. Berret, J.C. Castaing, F. Destremaut, J.B. Salmon, F. Cousin, J.P. Chapel, Influence of the Formulation Process in Electrostatic Assembly of Nanoparticles and Macromolecules in Aqueous Solution: The Interaction Pathway, J. Phys. Chem. C, 114 (2010) 16373-16381.
[51] D. Priftis, K. Megley, N. Laugel, M. Tirrell, Complex coacervation of poly(ethylene-imine)/polypeptide aqueous solutions: Thermodynamic and rheological characterization, J. Colloid Interface Sci., 398 (2013) 39-50.
[52] V. Mengarelli, L. Auvray, D. Pastre, M. Zeghal, Charge inversion, condensation and decondensation of DNA and polystyrene sulfonate by polyethylenimine, Eur. Phys. J. E, 34 (2011) 127.
[53] M. Miyake, K. Ogawa, E. Kokufuta, Light-Scattering Study of Polyelectrolyte Complex Formation between Anionic and Cationic Nanogels in an Aqueous Salt-Free System, Langmuir, 22 (2006) 7335 - 7341.
[54] J.P. Chapel, J.-F. Berret, Versatile electrostatic assembly of nanoparticles and polyelectrolytes: Coating, clustering and layer-by-layer processes, Curr. Opin. Colloid Interface Sci., 17 (2012) 97-105.
[55] A. Van Blaaderen, J. Van Geest, A. Vrij, Monodisperse colloidal silica spheres from tetraalkoxysilanes: Particle formation and growth mechanism, J. Colloid Interface Sci., 154 (1992) 481-501.
[56] N. Reinhardt, L. Adumeau, O. Lambert, S. Ravaine, S. Mornet, Quaternary Ammonium Groups Exposed at the Surface of Silica Nanoparticles Suitable for DNA Complexation in the Presence of Cationic Lipids, J. Phys. Chem. B, 119 (2015) 6401-6411.
[57] I. George, G. Naudin, S. Boland, S. Mornet, V. Contremoulins, K. Beugnon, L. Martinon, O. Lambert, A. Baeza-Squiban, Metallic oxide nanoparticle translocation across the human bronchial epithelial barrier, Nanoscale, 7 (2015) 4529-4544.
[58] S. Desset, O. Spalla, P. Lixon, B. Cabane, From powders to dispersions in water: Effect of adsorbed molecules on the redispersion of alumina particles, Langmuir, 17 (2001) 6408-6418.
[59] R.G. Gilbert, M. Hess, A.D. Jenkins, R.G. Jones, R. Kratochvil, R.F.T. Stepto, Dispersity in polymer science (IUPAC Recommendations 2009) (vol 81, pg 351, 2009), Pure and Applied Chemistry, 81 (2009) 779-779.
[60] J.-F. Berret, Stoichiometry of electrostatic complexes determined by light scattering, Macromolecules, 40 (2007) 4260-4266.
[61] J.-F. Berret, A. Sehgal, M. Morvan, O. Sandre, A. Vacher, M. Airiau, Stable oxide nanoparticle clusters obtained by complexation, J. Colloid Interface Sci., 303 (2006) 315-318.
[62] I. Jelesarov, H.R. Bosshard, Isothermal titration calorimetry and differential scanning calorimetry as complementary tools to investigate the energetics of biomolecular recognition, Journal of Molecular Recognition, 12 (1999) 3 - 18.
[63] E. Spruijt, A.H. Westphal, J.W. Borst, M.A.C. Stuart, J. van der Gucht, Binodal Compositions of Polyeletrolyte Complexes, Macromolecules, 43 (2010) 6476-6484.
[64] E. Spruijt, J. Sprakel, M. Lemmers, M.A.C. Stuart, J. van der Gucht, Relaxation Dynamics at Different Time Scales in Electrostatic Complexes: Time-Salt Superposition, Phys. Rev. Lett., 105 (2010).
[65] D. Priftis, R. Farina, M. Tirrell, Interfacial Energy of Polypeptide Complex Coacervates Measured via Capillary Adhesion, Langmuir, 28 (2012) 8721-8729.
[66] F. Bordi, C. Cametti, S. Sennato, M. Diociaiuti, Direct evidence of multicompartment aggregates in

polyelectrolyte-charged liposome complexes, Biophys. J., 91 (2006) 1513-1520.

[67] A.I. Petrov, A.A. Antipov, G.B. Sukhorukov, Base-acid equilibria in polyelectrolyte systems: From weak polyelectrolytes to interpolyelectrolyte complexes and multilayered polyelectrolyte shells, Macromolecules, 36 (2003) 10079-10086.